\documentstyle[11pt,iau185,twoside,epsf,psfrag]{article}

\begin{document}

\title{A comparison of the anelastic and subseismic approximations for
low-frequency stellar oscillations: an application to rapidly rotating
stars}

\author{B. Dintrans\altaffilmark{1}}
\affil{Nordic Institute for Theoretical Physics, Copenhagen, Denmark}
\affil{Observatoire Midi-Pyr\'en\'ees, 14 av. E. Belin, 31400 Toulouse, France}
\altaffiltext{1}{Supported by the European Commission under Marie-Curie
grant No.~HPMF-CT-1999-00411; email: boris.dintrans@ast.obs-mip.fr}

\author{M. Rieutord}
\affil{Observatoire Midi-Pyr\'en\'ees, 14 av. E. Belin, 31400 Toulouse, France}
\affil{Institut Universitaire de France}

\begin{abstract}
After showing that the anelastic approximation is better than the
subseismic one to filter out acoustic waves when studying low-frequency
stellar oscillations, we compute gravito-inertial modes of a
typical $\gamma$-Doradus star using this approximation. We show that
eigenmodes can be regular or singular, according to the possible focusing
towards attractors of the underlying characteristics. Consequences on the
oscillations spectrum are then discussed.  
\end{abstract}

 
\section{Introduction}

$\gamma$-Doradus stars are rapidly rotating variables for which the
second-order perturbative theory of Ledoux (1951) fails to reproduce
observations. Indeed, oscillations and rotation periods are both around
one day (Handler \& Krisciunas 1997) and eigenfrequencies should be
non-perturbatively computed.

In this paper, gravito-inertial modes of a typical $\gamma$-Dor star
are computed using the anelastic approximation. We first
compare the anelastic and subseismic approximations and show that the
anelastic one is the best to filter out acoustic waves from the infinite
system of differential equations (\S 2). Properties of gravito-inertial
oscillations are then recalled (\S 3) and the part played by the
characteristics trajectories is enlightened, before conclusions in \S 4.

\section{Comparison of the anelastic and subseismic approximations}

These two approximations have been compared in Dintrans \& Rieutord
(2001). Both approximations neglect perturbations in the gravitational
potential and assume that the Eulerian pressure fluctuations do not
contribute to Lagrangian ones. On the contrary, they differ on the form
of the equation of mass conservation as

\[
\hbox{Anelastic:} \quad \overrightarrow{\nabla} \cdot (\rho_0 \vec{v}) = 0, 
\quad \hbox{Subseismic:} \quad \overrightarrow{\nabla} \cdot 
\vec{v} = (g/c^2) v_r,
\]
where $\rho_0$, $g$ and $c^2$ respectively denote the equilibrium density,
gravity and square of the velocity of sound whereas $\vec{v}$ is the
velocity. 

Comparisons have been made using g-modes of the homogeneous and $n=3$
polytropes. Analytic expressions for the anelastic and subseismic
eigenfrequencies have been found for the homogeneous model whereas all the
eigenfrequencies have been computed numerically  for the $n=3$ polytrope.

Results are summarized in Table 1. They show that the anelastic
values are between 5 and 20 times more accurate than the corresponding
subseismic ones demonstrating that the anelastic approximation is best
to filter out acoustic waves when studying low-frequency stellar
oscillations.

\begin{table}
\caption{Complete dimensionless eigenfrequencies $\omega^2$ for some
g-modes of the $n=3$ and homogeneous (in parenthesis) polytropes,
with their associated anelastic $\omega^2_{\rm{anel}}$ and subseismic
$\omega^2_{\rm{sub}}$ counterparts.}
\tabcolsep=10pt
\begin{center}
\begin{tabular}{lccc}
\hline\\[-10pt] 
         & $\omega^2 \times 10^3$ & $\omega^2_{\rm{anel}} \times 10^3$ &
$\omega^2_{\rm{sub}} \times 10^3$ \\[2pt] 
\hline\\[-10pt]
g$_5$    & 5.76059 (-36.1367) & 6.03954 (-35.2941) & 6.27794 (-32.6087) \\
g$_{10}$ & 1.98396 (-12.2196) & 2.02646 (-12.1212) & 2.08345 (-11.5607) \\
g$_{20}$ & 0.59984 (-3.65629) & 0.60453 (-3.64741) & 0.61587 (-3.55239) \\
g$_{30}$ & 0.28631 (-1.73527) & 0.28744 (-1.73327) & 0.29154 (-1.70180)
\\[2pt]\hline
\end{tabular}
\end{center}
\end{table}

\section{Application to the long-period oscillations of a $\gamma$-Dor
type-star}

The anelastic equations need to be solved in a co-rotating frame; they
read (Dintrans \& Rieutord 2000, hereafter DR2000)

\begin{equation}
i \sigma \overrightarrow{\nabla} \times \left( \vec{v} - \frac{N^2}{\sigma^2}
\vec{v} \cdot \vec{e}_r \right) + \overrightarrow{\nabla} \times ( 2 
\overrightarrow{\Omega} \times \vec{v} ) = 0 \quad \hbox{and} \quad 
\overrightarrow{\nabla} \cdot (\rho_0 \vec{v}) = 0,
\label{eigenv}
\end{equation}
where we assumed a time-dependence of the form $\exp(i \sigma t)$ and $N$
denotes the Brunt-V\"ais\"al\"a frequency. After projecting on vectorial
spherical harmonics (see Dintrans, Rieutord \& Valdettaro 1999 for
numerical details), one obtains a generalized eigenvalue problem which
is solved using either the QZ algorithm (all of eigenvalues $\sigma$
are computed) or an iterative Arnoldi-Chebyshev solver (only desired
eigenvalues $\sigma_n$ are computed, with their associated eigenvectors
$\vec{v}_n$).

The perturbative theory has been tested by following a high-order
g-mode (period$\sim$0.3d) of a typical 1.5 M$_{\sun}$ $\gamma$-Dor
star as rotation was increased. As expected for this kind of long-period
mode, perturbative theories rapidly reach their limits since first-
and second-order relative errors are comparable to the observational
detection limit for rotation periods of about 8 and 3 days, respectively.

\begin{figure}
\psfrag{titre1}{\hspace{-1cm} (a): $\sigma/2\Omega=10$}
\psfrag{titre2}{\hspace{-1.1cm} (b): $\sigma/2\Omega=2$}
\psfrag{titx}{$s$}
\psfrag{tity}{$z$}
\centerline{\epsfsize=0.5\textwidth\epsfbox{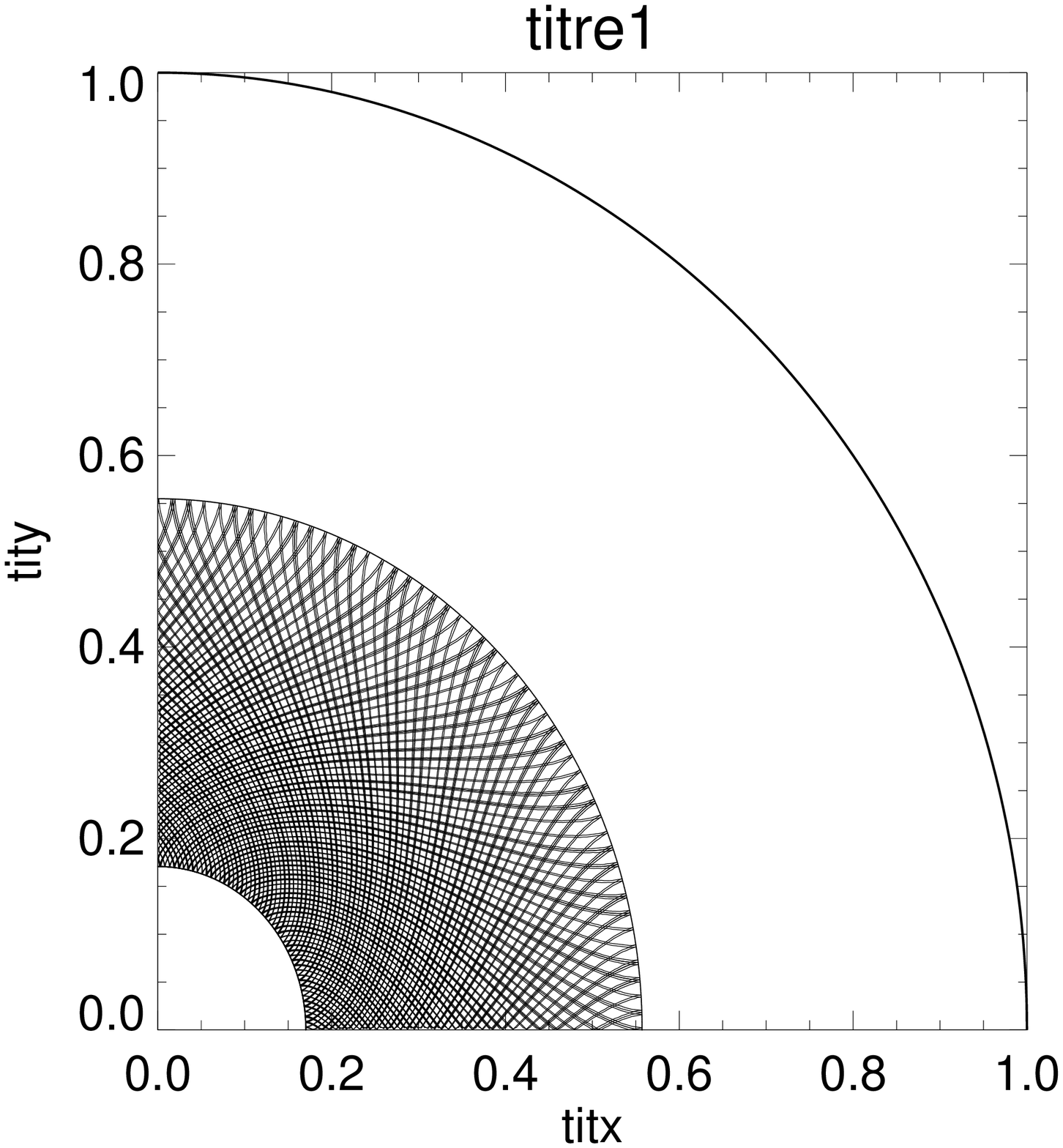}
\epsfsize=0.5\textwidth\epsfbox{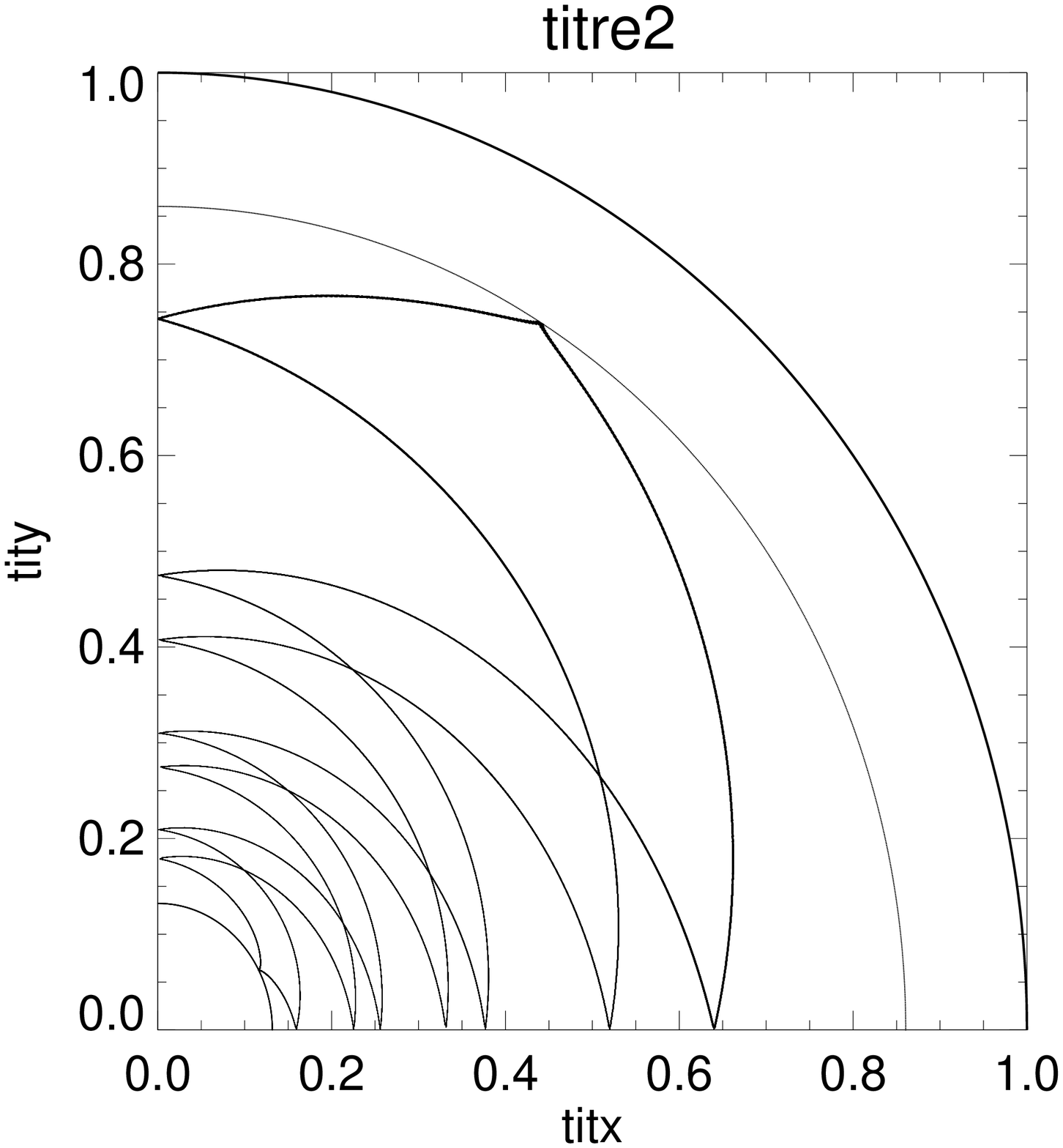}}
\caption{Integration of the characteristics equation (\ref{carac}) for
a star rotation period of one day and two mode frequencies $\sigma$. If
$\sigma$ is well above $2\Omega$ (a), characteristics fill the whole
hyperbolic domain whereas they can be focused along attractors when
$\sigma \sim 2\Omega$ (b). Three hundred reflections have been drawn in
both cases but we removed the first two hundred ones in plot (b) to
emphasize the final attractor.} 
\end{figure}

Then, by solving the non-perturbative problem (\ref{eigenv}) in the rapid
rotation r\'egime (rotation down to one day), we found the more and
more important part played by the characteristics trajectories as the
rotation increases. Characteristics, on which energy propagates, obey
the following differential equation ($s$ and $z$ being cylindrical
coordinates with $r^2=s^2+z^2$)

\begin{equation}
\frac{dz}{ds} = \frac{N^2 s z \pm r \sqrt{\sigma^2 N^2 s^2 + (4\Omega^2 -
\sigma^2)(\sigma^2 r^2-N^2 z^2)}}{\sigma^2 r^2 - N^2 z^2}.
\label{carac}
\end{equation}
Given a mode frequency $\sigma$ and rotation rate $\Omega$, this equation
can be integrated in the radiative zone of the star and we found that
the resulting web imposes the mode structure depending on the ratio
$\sigma/2\Omega$:

\begin{itemize}

\item if $\sigma \gg 2\Omega$, all characteristics trajectories are
ergodic (Fig.~1a) and the associated eigenmodes are {\it regular},
in the sense that the velocity field is smooth and square-integrable
(see Fig.~8a in DR2000).

\item if $\sigma \sim 2\Omega$ (the r\'egime of interest), characteristics
trajectories almost focus along attractors (Fig.~1b) and the associated
eigenmodes are {\it singular}, that is, velocity tends to infinity on
the attractor (see Fig.~8b in DR2000). Thus, without an efficient
diffusion process, these modes cannot exist.

\item if $\sigma < 2\Omega$, no global eigenmodes have been found, most
likely because of numerical difficulties.

\end{itemize}

\section{Conclusion}

First, we showed that the anelastic approximation is better than
the subseismic one to filter out acoustic waves from oscillations
equations. Second, we applied this result to investigate the low-frequency
oscillations of a rapidly rotating $\gamma$-Dor type-star.

We tested first- and second-order perturbative theories by following
a long-period g-mode while increasing rotation. We showed that a
non-perturbative approach is necessary for rotation periods below 3
days. Then, a computation of eigenmodes allowed us to reach the
one-day rotation r\'egime where we found that modes part into regular
and singular ones as the associated characteristic trajectories are
ergodic or focused along an attractor. It has important consequences on
the oscillations spectrum since frequency subintervals corresponding to
attractors are lacking eigenvalues when dissipative effects are neglected.

\section*{Discussion}
\parindent=0pt

{\it R. Townsend~:} What advantage does the anelastic approximation give,
if you must still use an expansion in spherical harmonics ?

\medskip

{\it B. Dintrans~:} We always need to expand the angular part of
perturbations onto spherical harmonics. But adding rotation through
Coriolis' force leads to an infinite system of coupled differential
equations. The usefulness of the anelastic approximation is precisely
to decrease the system size by filtering out acoustic waves, which makes
numerics more efficient.

\bigskip

{\it R. Townsend~:} What is the physical interpretation of the singular
modes which you find ?

\medskip

{\it B. Dintrans~:} These modes are due to the
focusing of the characteristics, associated with the inviscid problem,
towards periodic orbits (attractors); this process is therefore coming
from the ill-posedness of the associated mathematical problem. Physically,
you may think to a wave packet which, following the characteristics, will
bounce on the boundaries. After each cycle, its wavelength is reduced
by a factor (which depends on frequency) larger than unity.  After an
infinite time, the wavelength is zero and the singularity is born.

\end{document}